\documentclass[12p]{article}
\usepackage{eso-pic} 
\usepackage{latexsym}
\usepackage{color}
\usepackage{amsmath}
\usepackage{epsfig}
\usepackage{hyperref}
 \usepackage{subfigure}
 \usepackage{dcolumn}
   \usepackage{threeparttable}

\newcommand{\ortala}[1]{\begin{center}#1\end{center}}

\newcommand{\integ}[3]{{{\underset{#1 }{\overset{#2}{\displaystyle\int}}}#3}}

\newcommand{\summ}[3]{{{\underset{#1 }{\overset{#2}{\displaystyle\sum}}}#3}}

\newcommand{\re}[1]{(\ref{#1})}

\newcommand{\eq}[2]{\begin{equation}\label{#1}  #2\end{equation}}

\newcommand{\paran}[1]{\left(#1\right)}

\newcommand{\sch}[1]{Schrodinger}

\newcommand{\komb}[2]{\paran{\begin{array}{c} #1 \\ #2 \end{array}}}

\linethickness{0.6pt}

\begin{document}

\ortala{\textbf{Effects of the randomly distributed magnetic field on the phase diagrams of the Ising Nanowire I: discrete distributions}}

\ortala{\textbf{\"Umit Ak\i nc\i \footnote{umit.akinci@deu.edu.tr}}}

\ortala{\textit{Department of Physics, Dokuz Eyl\"ul University,
TR-35160 Izmir, Turkey}}

\section{Abstract}

The effect of the random magnetic field distribution on the phase
diagrams and ground state magnetizations of the Ising nanowire is
investigated with effective field theory with correlations. Trimodal
distribution chosen as a random magnetic field distribution. The
variation of the phase diagrams with that distribution parameters
obtained and some interesting results found such as reentrant
behavior. Also for the trimodal distribution, ground state
magnetizations for different distribution parameters determined
which can be regarded as separate partially ordered phases of the
system.
Keywords: \textbf{Ising Nanowire; random magnetic field; trimodal distribution}

\section{Introduction}\label{introduction}

Recently there has been growing interest both theoretically and
experimentally in the magnetic nanomaterials such as nanoparticles,
nanorods, nanotubes and nanowires. Nowadays, fabrication of these
nanomaterials no longer difficult, since development of the
experimental techniques permit us making materials with a few atoms.
For instance acicular magnetic nano elements were fabricated
\cite{ref1,ref2,ref3} and  magnetization of the nanomaterial has
been measured \cite{ref4}. In general nanoparticle systems can be
used as sensors \cite{ref60} and they can be used for making
permanent magnets \cite{ref62}, beside some medical applications
\cite{ref61}. In particular, magnetic nanowires and nanotubes have
many applications to a nanotechnology \cite{ref53,ref54}. Nanowires
can be used as an ultrahigh density magnetic recording media
\cite{ref36,ref37,ref59}   and they have potential applications in
biotechnology \cite{ref42,ref43}, such as Ni nanowires can be used
for bio seperation \cite{ref69,ref70}.

In a nanometer scale for the magnetic properties of these materials,
some unexpected properties appear and in a nanotechnology these
properties allow us to fabrication materials for various purposes,
since behaviors of these finite materials different than bulk
counterparts. In addition, magnetic properties and magnetic phase
transition characteristics are highly depends on the size and the
dimensionality of the material.

Mean field approximation (MFA), effective field theory (EFT) and
Monte Carlo (MC) simulation are most common used theoretical methods
for determining the magnetic properties of these materials.  For
instance,  nanoparticles investigated by EFT with correlations
\cite{ref44}, MFA and MC \cite{ref45}. The phase diagrams and the
magnetizations of the nanoparticle described by the transverse Ising
model investigated with using MFA and EFT \cite{ref34,ref35},
compensation temperature of the nano particle \cite{ref67} and
magnetic properties of the nanocube with MC \cite{ref33} are among
these studies.

Another method namely Variational cumulant expansion (VCE) based on
expanding free energy in terms of the action up to $m^{th}$ order,
has been applied to the magnetic superlattices \cite{ref10} and
ferromagnetic nanoparticles \cite{ref11,ref12}. The first order
expansion within this method gives the results of the MFA.

Various nano structures such as $FePt$ and $Fe_3 O_4$ nanotubes
\cite{ref51} can be modeled by core-shell models and these models
can be solved also by MFA, EFT and MC. The phase diagrams and
magnetizations of the transverse Ising nanowire is treated within
MFA and EFT \cite{ref38,ref39}, the effect of the surface dilution
on the magnetic properties of the cylindrical Ising nanowire and
nanotube has been studied \cite{ref40,ref58}, the magnetic
propeties of nanotubes of different diameters, using armchair or
zigzag edges has been investigated with MC \cite{ref41}, initial
susceptibility of the Ising nanotube and nanowire within the EFT
with correlations \cite{ref46,ref47} and the compensation
temperature which appears negative core-shell coupling is
investigated by EFT for nanowire and nanotube \cite{ref52}. There
has been also works deal with hysteresis characteristics of the
cylindrical Ising nanowire \cite{ref57,ref71}. Beside these, higher
spin nanowire or nanotube has been investigated also, e.g. spin - 1
nanotube \cite{ref63} and nanowire \cite{ref64}, mixed spin - $3/2,1$ core
shell structured nanoparticle \cite{ref65}, mixed spin - $1/2,1$
nanotube \cite{ref66}.

On the other hand, as far as we know, there is less attention to
quenched randomness effects of these systems except the site
dilution. However, including quenched randomness or disorder effects
to these systems may induce some beneficial results. For this
purpose we investigate the effects of the random magnetic field
distributions on the phase diagrams of the Ising nanowire within
this work. As stated in \cite{ref40} the phase diagrams of the
nanotube and nanowire are similar qualitatively, then investigation
of the effect of the random magnetic field distribution on the
nanowire will give hints about the effect of the same distribution
on the phase diagrams of the nanotube.

The Ising model in a quenched random field (RFIM) has been studied
over three decades. The model which is actually based on the local
fields acting on the lattice sites which are taken to be random
according to a given probability distribution was introduced for the
first time by Larkin \cite{refs1} for superconductors and later
generalized by Imry and Ma \cite{refs2}.   Beside the equality
between the diluted antifferromagnets and ferromagnets with random
field distribution \cite{refs3,refs4}, the importance of the random
field distribution on these systems comes from  the fact that,
random distribution of the magnetic field changes drastically phase
diagrams of the system and then magnetic properties.

The paper is organized as follows: In Sec. \ref{formulation} we
briefly present the model and  formulation. The results and
discussions are presented in Sec. \ref{results}, and finally Sec.
\ref{conclusion} contains our conclusions.

\section{Model and Formulation}\label{formulation}

\begin{figure}[h]\begin{center}
\epsfig{file=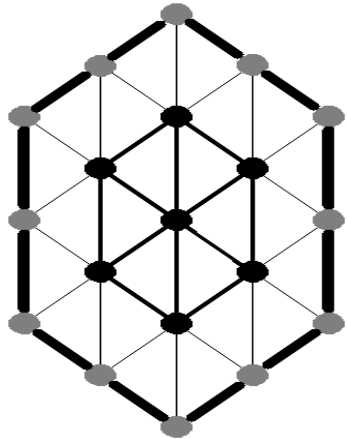, width=5cm,height=6cm}
\end{center}
\caption{Schematic representation of a cylindrical nanowire (top view). The gray/black circles represents of the surface/shell magnetic atoms, respectively}
\label{sek1}\end{figure}

We consider a nanowire which has geometry shown in Fig. \ref{sek1}.
The Hamiltonian of the nanowire is given by

\eq{denk1}{\mathcal{H}=-J_{1}\summ{<i,j>}{}{s_is_j}-J_{2}\summ{<m,n>}{}{s_ms_n}-J_3\summ{<i,m>}{}{s_is_m}-\summ{i}{}{H_is_i}-\summ{m}{}{H_ms_m}}
 where $s_i$ is the $z$ component of the spin at a lattice site $i$ and it takes the values $s_i=\pm 1$ for the spin-1/2 system, $J_1$ and $J_2$ are the exchange
interactions between spins which are located at the core and shell,
respectively, and $J_3$ is the exchange interaction between the core
and shell spins which are nearest neighbor to each other. $H_i$ and
$H_m$ are the external longitudinal magnetic fields at the lattice
sites $i$ and $m$ respectively. Magnetic fields are distributed to
lattice sites by a given probability distribution. The first three
summations in Eq. \re{denk1} is over the nearest-neighbor pairs of
spins, and the other summations are over all the lattice sites.

This work deals with the following discrete magnetic field distribution, namely trimodal distribution
\eq{denk2}{P\paran{H_i}=p\delta\paran{H_i}+\frac{1-p}{2}\left[\delta\paran{H_i-H_0}+\delta\paran{H_i+H_0}\right]
} which covers a bimodal distribution for $p=0$ and reduces to the system with zero magnetic field (pure system) for $p=1$. According to the distribution given in Eq. \re{denk2}, $p$ percentage of the lattice sites are subjected to a magnetic field $H_i=0$, while half of the remaining sites are under the influence of a field $H_i=H_0$  whereas the field $H_i=-H_0$ acts on the remaining sites.

We have to separate four representative spins according to their interactions with other spins. Their magnetizations ($m_i,i=1,2,3,4$) can be given by usual EFT equations which are obtained by differential operator technique and decoupling approximation (DA) \cite{refs5,refs6},
\eq{denk3}{\begin{array}{lcl}
m_1&=&\left[A_1+m_1B_1\right]^4\left[A_3+m_2B_3\right]\left[A_3+m_3B_3\right]^2\left[A_1+m_4B_1\right]\\
m_2&=&\left[A_3+m_1B_3\right]\left[A_2+m_2B_2\right]^2\left[A_2+m_3B_2\right]^2\\
m_3&=&\left[A_3+m_1B_3\right]^2\left[A_2+m_2B_2\right]^2\left[A_2+m_3B_2\right]^2\\
m_4&=&\left[A_1+m_1B_1\right]^6\left[A_1+m_4B_1\right]^2\\
\end{array}}
Here $m_1,m_4$ are the magnetizations of the two different representative sites in the core and  $m_2,m_3$ are the magnetizations of the two different representative sites in the shell. The coefficients in Eq. \re{denk3} are
\eq{denk4}{\begin{array}{lcl}
A_n&=&\cosh\paran{J_{n}\nabla}F\paran{x}|_{x=0}\\
B_n&=&\sinh\paran{J_{n}\nabla}F\paran{x}|_{x=0}\\
\end{array}} where $\nabla$ is the usual differential operator in the differential operator technique  and $n=1,2,3$. The
function is defined by
\eq{denk5}{F\paran{x}=\integ{}{}{}dH_iP\paran{H_i}f\paran{H_i,x}}
where \eq{denk6}{f\paran{H_i,x}=\tanh\paran{\beta x + \beta H_i}.}
In Eq. \re{denk6}, $\beta=1/(k_B T)$ where $k_B$ is Boltzmann
constant and $T$ is the temperature. The effect of the exponential
differential operator to a function is given by
\eq{denk7}{\exp{\paran{a\nabla}}F\paran{x}=F\paran{x+a}} with any
constant  $a$. DA will give the results of the Zernike approximation
\cite{refs7} for this system.

With the help of the Binomial expansion, Eq. \re{denk3} can be written in the form
\eq{denk8}{\begin{array}{lcl}
m_1&=&\summ{i=0}{4}{}\summ{j=0}{1}{}\summ{k=0}{2}{}\summ{l=0}{1}{}K_1\paran{i,j,k,l}m_1^im_2^jm_3^k m_4^l\\
m_2&=&\summ{i=0}{1}{}\summ{j=0}{2}{}\summ{k=0}{2}{}K_2\paran{i,j,k}m_1^i m _2^j m_3^k\\
m_3&=&\summ{i=0}{2}{}\summ{j=0}{2}{}\summ{k=0}{2}{}K_3\paran{i,j,k}m_1^i m_2^j m_3^k\\
m_4&=&\summ{i=0}{6}{}\summ{l=0}{2}{}K_4\paran{i,l}m_1^i m_4^l\\
\end{array}} where

\eq{denk9}{\begin{array}{lcl}
K_1\paran{i,j,k,l}&=&\komb{4}{i}\komb{1}{j}\komb{2}{k}\komb{1}{l}A_1^{5-i-l}A_3^{3-j-k}B_1^{i+l}B_3^{j+k}\\
K_2\paran{i,j,k}&=&\komb{1}{i}\komb{2}{j}\komb{2}{k}A_2^{4-j-k}A_3^{1-i}B_2^{j+k}B_3^{i}\\
K_3\paran{i,j,k}&=&\komb{2}{i}\komb{2}{j}\komb{2}{k}A_2^{4-j-k}A_3^{2-i}B_2^{j+k}B_3^{i}\\
K_4\paran{i,l}&=&\komb{6}{i}\komb{2}{l}A_1^{8-i-l}B_1^{i+l}\\
\end{array}}
These coefficients can be calculated from the definitions given in Eqs. \re{denk4} and \re{denk7}.

For a given Hamiltonian and field distribution parameters, by determining the coefficients  from Eq. \re{denk9} we can obtain a system of coupled non linear equations from Eq. \re{denk8}, and by solving this system we can get the magnetizations $m_i,i=1,2,3,4$. The magnetization of the core $(m_c)$ and shell $(m_s)$ of nanowire, as well as the total magnetization $(m_T)$ can be calculated via
\eq{denk10}{m_c=\frac{1}{7}\paran{6m_1+m_4}, \quad
m_s=\frac{1}{12}\paran{6m_2+6m_3}, \quad
m_T=\frac{1}{19}\paran{6m_1+6m_2+6m_3+m_4}}

Since in the vicinity of the critical point all magnetizations are close to zero, we can obtain another coupled  equation system for determining this critical point by linearizing the equation system given in  Eq. \re{denk8}, i.e.
\eq{denk11}{A.m=0}
where
\eq{denk12}{A=\left(
\begin{array}{cccc}
K_1(1,0,0,0)&K_1(0,1,0,0)&K_1(0,0,1,0)&K_1(0,0,0,1)\\
K_2(1,0,0)&K_2(0,1,0)&K_2(0,0,1)&0\\
K_3(1,0,0)&K_3(0,1,0)&K_3(0,0,1)&0\\
K_4(1,0)&0&0&K_4(0,1)\\
\end{array}
\right) }

\eq{denk13}{m=\left(
\begin{array}{c}
m_1\\
m_2\\
m_3\\
m_4\\
\end{array}
\right)}

Critical temperature can be determined from $det(A)=0$. As discussed in \cite{ref47}, the matrix $A$ given in Eq. \re{denk12} is invariant under the transformation $J_3\rightarrow -J_3$ then we can conclude that the system with ferromagnetic ($J_3>0$) core-shell interaction has the same critical temperature as that of the system with anti-ferromagnetic ($J_3<0$) core-shell interaction (with the same $|J_3|$) for certain Hamiltonian and magnetic field distribution parameters. Although this discussion has been made for the system with zero magnetic field in \cite{ref47}, this conclusion is also valid for this system, because of the symmetry of the magnetic field distribution. The equation $det(A)=0$ is invariant under the transformation $J_3\rightarrow -J_3$ for the nanowire with trimodal magnetic field distribution.

\section{Results and Discussion}\label{results}

In this work we discuss the effect of the random magnetic field distribution on the phase diagrams of the system. Since the phase diagrams are the same for the $J_3>0$ and $J_3<0$ with the same $|J_3|$, we focus ourselves on the case $J_3>0$, i.e ferromagnetic core-shell interaction. We use the scaled interactions as
\eq{denk14}{J_1=J, \quad r_n=\frac{J_n}{J}, \quad n=2,3.}
Let us start with bimodal magnetic field distribution.

\subsection{Bimodal Distribution}

This distribution is given by Eq. \re{denk2} with $p=0$. It is a
well known fact that, in this distribution, increasing $H_0$ values
drag the system to the disordered phase, due to the random
distribution of the magnetic fields $\pm H_0$ on the lattice sites.
On the other hand, the interactions $J_n,(n=1,2,3)$  enforce the
system to stay in the ordered phase. Thus a competition takes place
between these two factors, in addition to the thermal agitations.

\begin{figure}[h]\begin{center}
\epsfig{file=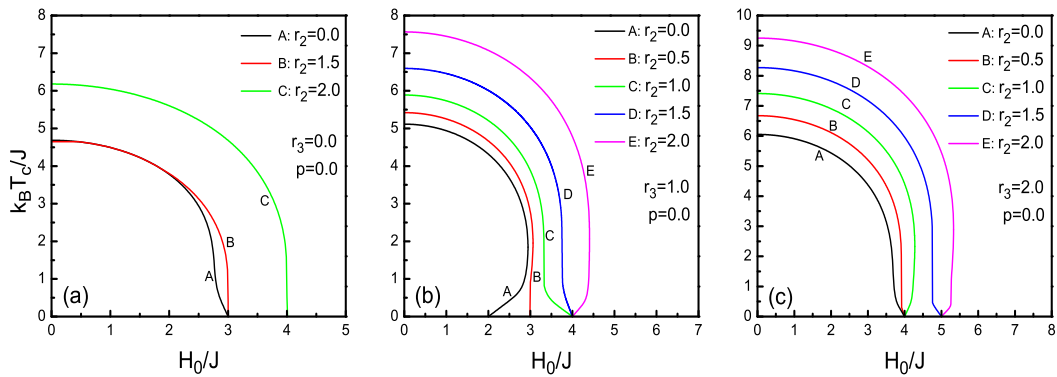, width=14cm,height=5cm}
\end{center}
\caption{The phase diagrams of the nanowire  with bimodal random
field distribution in the ($k_BT_c/J, H_0/J)$ plane for different
$r_2,r_3$ values.} \label{sek2}\end{figure}

\begin{figure}[h]\begin{center}
\epsfig{file=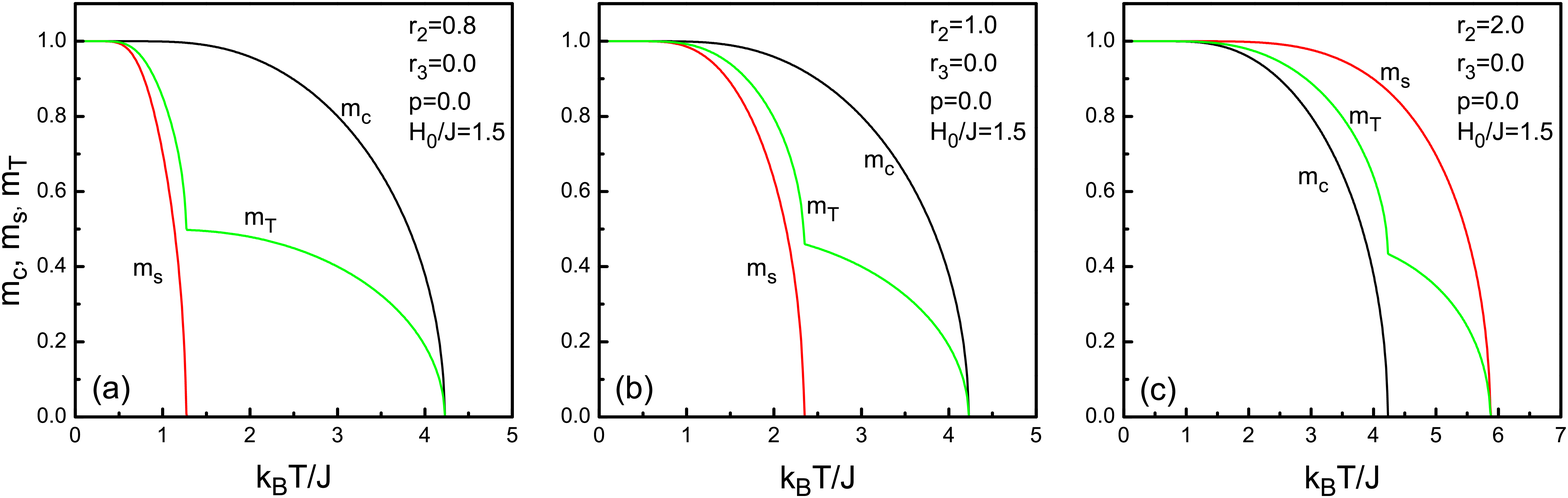, width=14cm,height=4cm}
\end{center}
\caption{Variation of the magnetization with temperature for some
selected $r_2$ parameter value of the nanowire with bimodal random
field distribution. Fixed parameter values  are
$r_3=0.0,H_0/J=1.5$.} \label{sek3}\end{figure}

The phase diagrams in $(k_BT_c/J, H_0/J)$ plane can be seen in Fig.
\ref{sek2} for different $r_2$ and $r_3$ values. We can see from Fig.
\ref{sek2} that the general effect of increasing $H_0$ is to decrease
the critical temperature, as expected. Besides, another expected
result is that the higher $r_2$ or $r_3$ values has the higher $T_c$
values at a certain $H_0$ value which can be seen in Fig. \ref{sek2}.
But we can see  in Fig. \ref{sek2}(a) that, the latter
observation does not hold for $r_3=0$ case. In the absence of the
interface interaction between the core and shell, the phase diagrams
in $(k_BT_c/J, H_0/J)$ plane are the same for all $r_2\le r_2^\star$
values. In order to clarify this situation, we plot the curves of
the core magnetization, shell magnetization and the total
magnetization of the system in Fig. \ref{sek3}. As seen in Fig.
\ref{sek3}, shell magnetization falls to zero at a temperature lower
than that of the core for the $r_2=0.8,1.0$. For these values, the
core determines the critical temperature of the system. But for the
value of $r_2=2.0$, the critical temperature of the shell is higher
than the critical temperature of the core. Then we can say that
shell of the nanowire has a lower critical temperature than that of
the core, up to a certain $r_2=r_2^\star$ value, after that
$r_2^\star$ value, the critical temperature of the shell becomes
higher than the critical temperature of the core i.e. the shell
begins to determine the critical temperature of the system. This
means that for the values that satisfies $r_2<r_2^\star$, the phase
diagram of the system does not affected by the changing $r_2$.
Although the $r_2=1.5$ curve in Fig. \ref{sek2}(a) is
different from other $r_2$ curves corresponding to $r_2<1.5$ for the
higher $H_0/J$ (which are close to the $H_0/J=3.0$) values, we can
say that this $r_2^\star$ value satisfies $r_2>1.0$. This fact can
be explained by taking a look at the geometry of the system. In the
absence of the interaction between the core and shell, both of the
two distinct shell spins have four nearest neighbors (which have
magnetizations labeled as $m_2,m_3$) and the two distinct core spins
have five and eight nearest neighbors, respectively (which have
magnetizations labeled as $m_1,m_4$). Thus we can say that for equal
exchange interactions of the core and shell ($r_2=1$), the core has
higher critical temperature than that of the shell. Consequently,
$r_2$ value that providing the equality of the critical temperatures
of the core and shell has to be $r_2>1.0$. In summary, we can
conclude that, for the $r_3=0$ phase diagrams in $(k_BT_c/J, H_0/J)$
plane, all diagrams are the same up to a certain $r_2^\star>1.0$
value, and when $r_2=r_2^\star$, the diagrams begin to separate from
the diagrams which are for the $r_2<r_2^\star$ in the higher $H_0/J$
region. Furthermore, as $r_2$ grows then the phase diagrams become
completely separate from the diagrams for the $r_2<r_2^\star$.

We can see from Fig. \ref{sek2} that, increasing $r_2$ and $r_3$
values make the ferromagnetic region wider in ($k_BT_c/J, H_0/J$)
plane, since increasing $r_2$ and $r_3$ values rises the absolute
value of the lattice energy coming from the spin-spin interaction,
which must be overcomed by both thermal agitations and random
magnetic fields for the phase transition from an ordered phase to a
disordered one.



\begin{figure}[h]\begin{center}
\epsfig{file=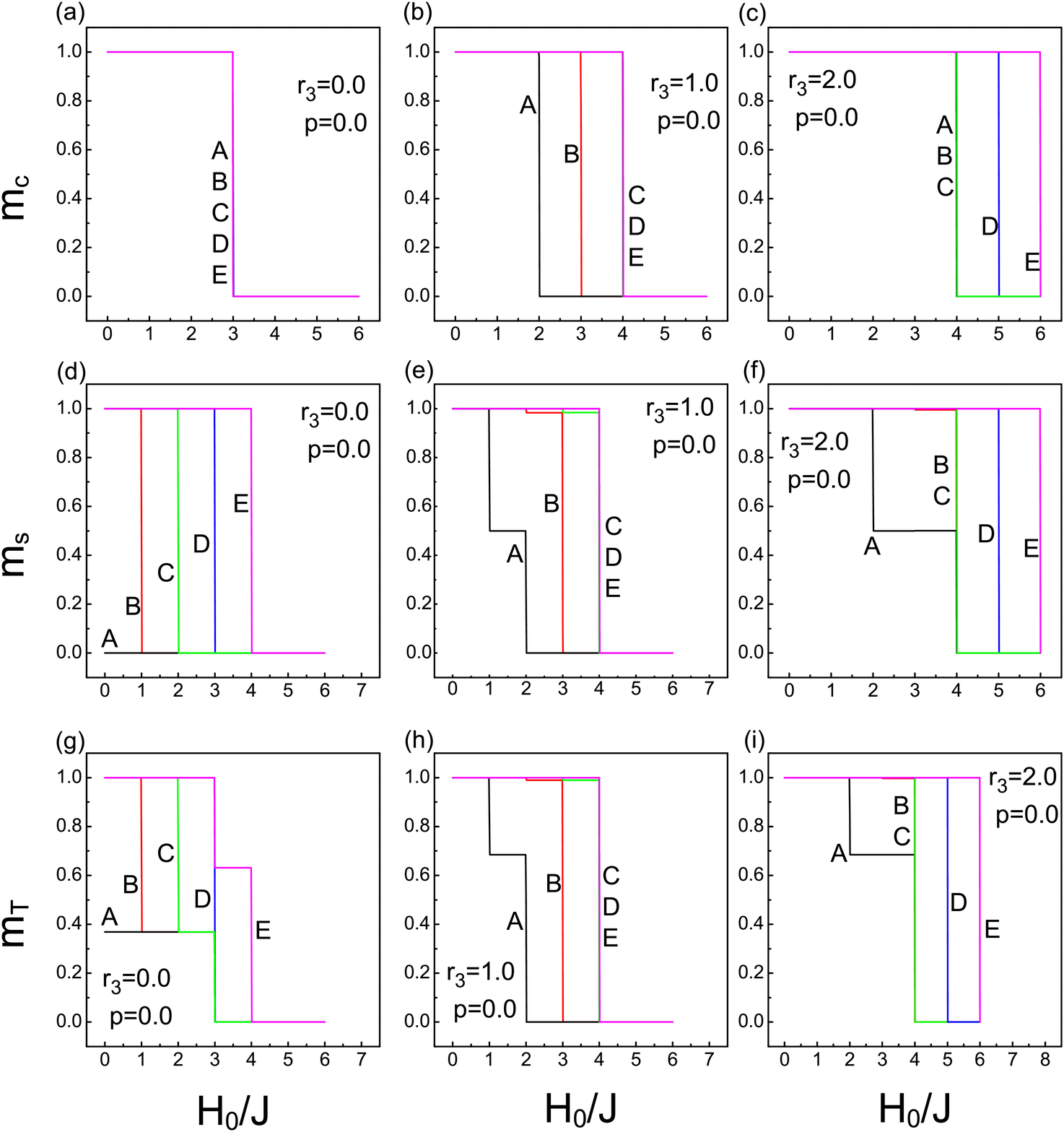, width=14cm}
\end{center}
\caption{Variation of the ground state magnetizations with $H_0/J$
which corresponds to the phase diagrams seen in Fig. \ref{sek2} for
the system with bimodal magnetic field distribution. Each
magnetization calculated at a temperature $k_BT/J=0.001$ which can
be regarded as  ground state. The letters which are close to the
related curve represent the different $r_2$ values as (A) $r_2=0.0$,
(B) $r_2=0.5$, (C) $r_2=1.0$, (D) $r_2=1.5$, (E) $r_2=2.0$}
\label{sek4}\end{figure}



\begin{figure}[h]\begin{center}
\epsfig{file=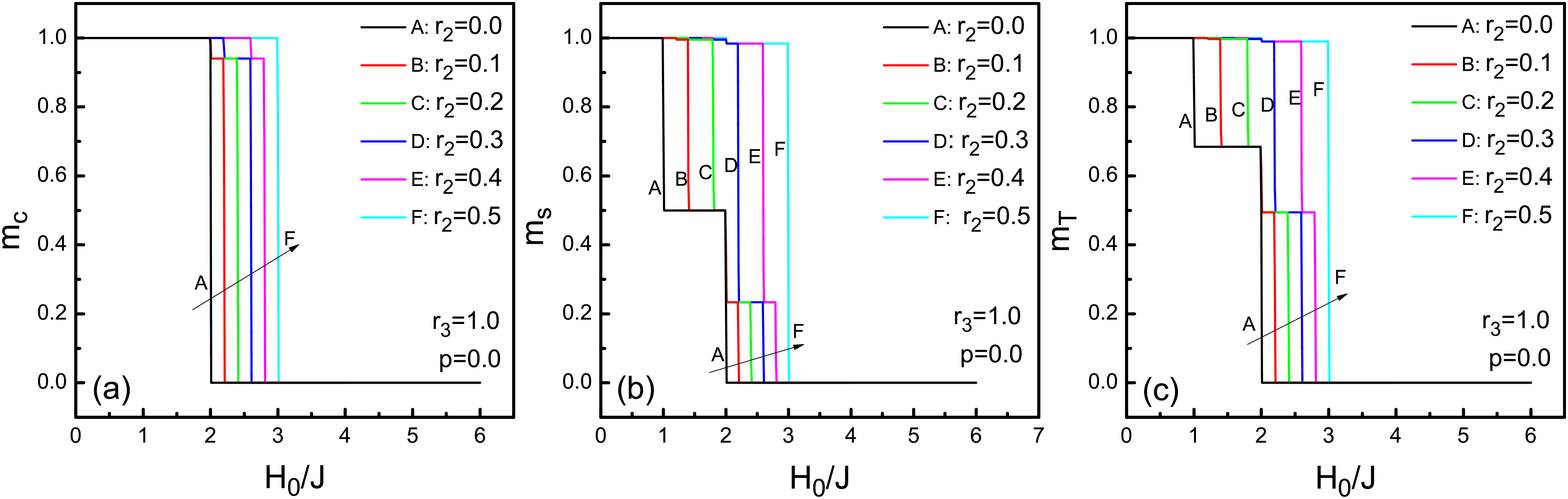, width=14cm}
\end{center}
\caption{Variation of the ground state magnetizations for selected
values of $r_2$ for the system with bimodal magnetic field
distribution. Each magnetization calculated at a temperature
$k_BT/J=0.001$ which can be regarded as  ground state and for the
parameter value $r_3=1.0$.} \label{sek5}\end{figure}

Now the question is: what is the magnetization behavior for that
different $r_2,r_3$ and $H_0$ values for bimodal magnetic field
distribution. Since the magnetization behavior with temperature is
well known for the nanowire without any random magnetic field and
due to the number of parameters in this problem and also the restrictions
on the length of the paper, we restrict ourselves to this question:
what is the ground state magnetization behaviors with $H_0$ form different $r_2,r_3$ values? We can
see the answer of the question from Fig. \ref{sek4} which is the
variation of the core ground state magnetization($m_c^{g}$), shell
ground state magnetization ($m_s^g$) and the total ground state
magnetization ($m_T^g$) (where the superscript $g$ stands for the
ground state)  with $H_0$. This figure
corresponds to the phase diagrams given in Fig. \ref{sek2} and the
temperature is chosen as $k_BT/J=0.001$ which can be regarded as
ground state for the system. As seen in Fig. \ref{sek4}(g), in the
absence of the core-shell interaction (i.e. $r_3=0$), two more
ground state magnetization appears apart from the magnetizations
$m_T^g=0.0$ and $m_T^g=1.0$ which are correspond to the disordered
state and ordered state, respectively. These two more ground state
magnetization values are $m_T^g=0.3684$ and $m_T^g= 0.6316$ and can
be regarded as two different partially ordered phases of the
nanowire under the bimodal magnetic field distribution. Since the
core has the ground state magnetization $m_c^g=1.0$ regardless of
the value of $r_2$ up to the value of $H_0/J=3.0$ (
Fig. \ref{sek4}(a)), that partially ordered phases seen in Fig. \ref{sek4}(g) has to come from the mismatch in
magnetization behaviors with $H_0$ in the core and shell. It can be
seen from Eq. \re{denk10} that the core contributes to the total
magnetization with the ratio $7/19$ while  this ratio for the shell
is $12/19$. Because of the difference of the regions which are core
and shell in ordered and disordered phases with $H_0$, this two
partially ordered phases appears in the regions seen in Fig. \ref{sek4}(g). For example, for the $r_3=0.0,r_2=2.0$, the
core has ground state magnetization $m_c^g=1.0$ up to $H_0/J=3.0$
and the shell has ground state magnetization $m_s^g=1.0$ up to
$H_0/J=4.0$ then a partially ordered phase $m_T^g=0.6316=12/19$
appears in the intersection of this two region which is
$3.0<H_0/J<4.0$.

The presence of the interaction between the core and shell makes
this situation slightly different. When $r_3\ne 0.0$ then the system
becomes whole instead of separate core and shell as in the case
$r_3=0.0$. Then the core and the shell can develop partially ordered
phases separately, thus new partially ordered phases can appear in
the system. As seen in  Fig. \ref{sek4}(e), shell
can develop two new partially ordered phases (other than phases that
corresponds to ordered $m_s^g=1.0$ and disordered $m_s^g=0.0$
phases) which have magnetizations $m_s^g=0.5$ ($r_2=0$) and $m_s^g=
0.9835$ ($r_2=0.5,1.0$) due to the random field and presence of the
interaction with the core. This makes the ground state magnetization
of the system as in Fig. \ref{sek4}(h) with
the same reasoning explained just above, with two partially ordered
phases namely $m_T^g= 0.6842$ ($r_2=0$) and $m_T^g=0.9896$
($r_2=0.5,1.0$). The increasing interaction strength between the
core and shell increases the magnetization of that two partially
ordered phases that has magnetization near to $1.0$ of the shell and
then the ground state magnetizations of the system become as seen in Fig. \ref{sek4}(i).

More detailed investigation on that partially ordered phases can
show that, indeed a few other partially ordered phases exist in the
shell even in the core. This fact can be seen in
Fig. \ref{sek5}. As in Fig. \ref{sek5}(a) for
$r_3=1.0$, with little $r_2$ (e.g. $r_2=0.3$), rising $H_0$ can
induce a partially ordered phase in the core which disappears with
rising $r_2$ (e.g. $r_2=0.5$). The same thing also be valid for the
shell as seen in Fig. \ref{sek5}(b)  and thus whole
system which can be seen in Fig. \ref{sek5}(c).

\subsection{Trimodal Distribution}

\begin{figure}[h]\begin{center}

\epsfig{file=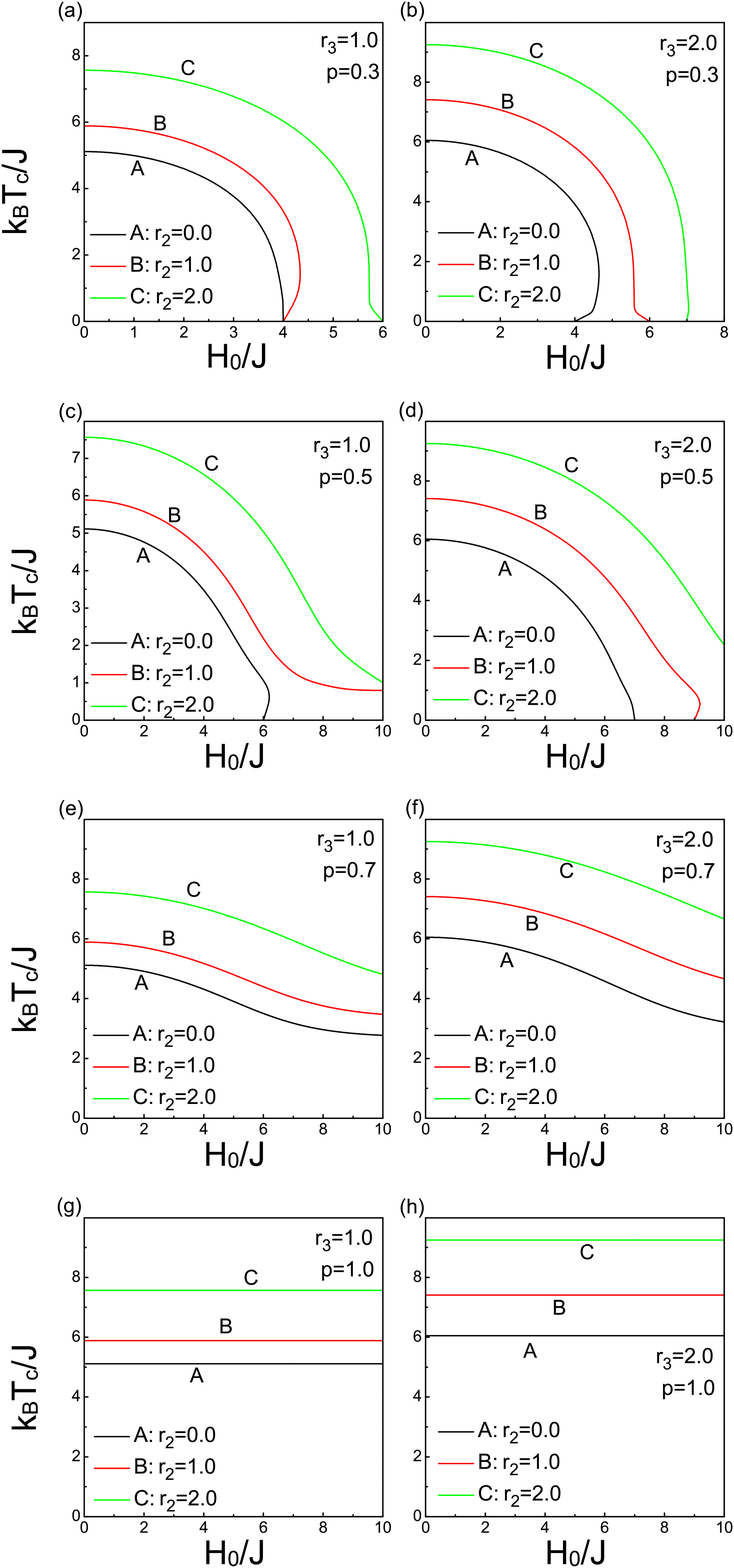, width=10cm}

\end{center}
\caption{The phase diagrams of the nanowire  with trimodal random field distribution in the $(k_BT_c/J, H_0/J)$ plane for different $r_2,r_3,p$ values}
\label{sek6}\end{figure}

Let us investigate the phase diagrams in the ($k_BT_c/J, H_0/J$)
plane for the distribution which is given by Eq. \re{denk2} for the
$p\ne 0$. We can see from the random magnetic field distribution
that, when $p$ approaches to $p=1$ from the value $p=0$, the phase
diagrams given in Fig. \ref{sek2} have to evolve to the phase
diagrams of the system with zero magnetic field, i.e. parallel lines
to the $H_0/J$ axes in that plane. This evolution can be seen in
Fig. \ref{sek6}.  We can see from the Fig. \ref{sek6} that (e.g. $r_2=1.0,r_3=1.0$ curve in order of Fig. \ref{sek6} (a),(c),(e),(g)) , rising $p$
from the $p=0$ to the $p=1$ first opens phase diagrams on the right hand
sides of them, then creates a tail on that side which extends to the
high magnetic fields then rises that tail to the level of the pure
system's critical temperature.  The same effect can be also performed
by the rising $r_2$ (or $r_3$) in the trimodal distribution for little $p$ values, since that
effect results from the competition between the spin-spin
interaction which tends system to ordered phase and randomly
distributed magnetic field. This can be seen for instance in Fig. \ref{sek6}(c). We also note that, during this
evolution process a phenomena called reentrant can be seen in the diagrams.

\begin{figure}[h]\begin{center}
\epsfig{file=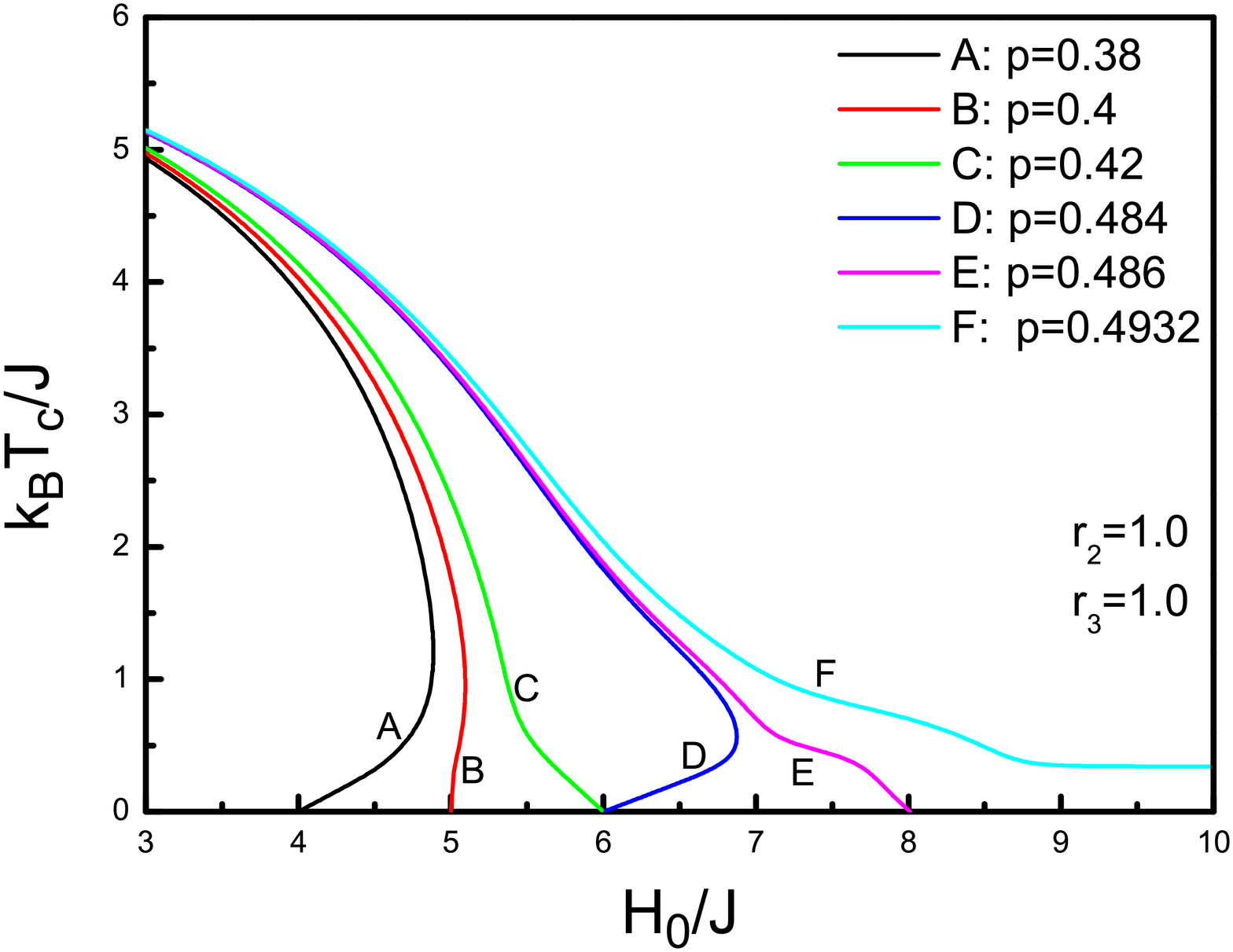, width=8cm,height=8cm}
\end{center}
\caption{The phase diagrams of the nanowire  with trimodal random field distribution in the $(k_BT_c/J, H_0/J)$ plane for different $p$ values. Fixed values of the diagrams are $r_2=1.0$ and $r_3=1.0$.}
\label{sek7}\end{figure}

The details of the evolution process of the phase diagram with $p$ in
the ($k_BT_c/J, H_0/J$) plane for fixed $r_2=1.0,r_3=1.0$ values can be seen in Fig. \ref{sek7}. We can see from
Fig. \ref{sek7} that, at zero temperature the transitions from the ordered phase to a disordered one occurs at integer $H_0/J$ values.

\begin{figure}[h]\begin{center}
\epsfig{file=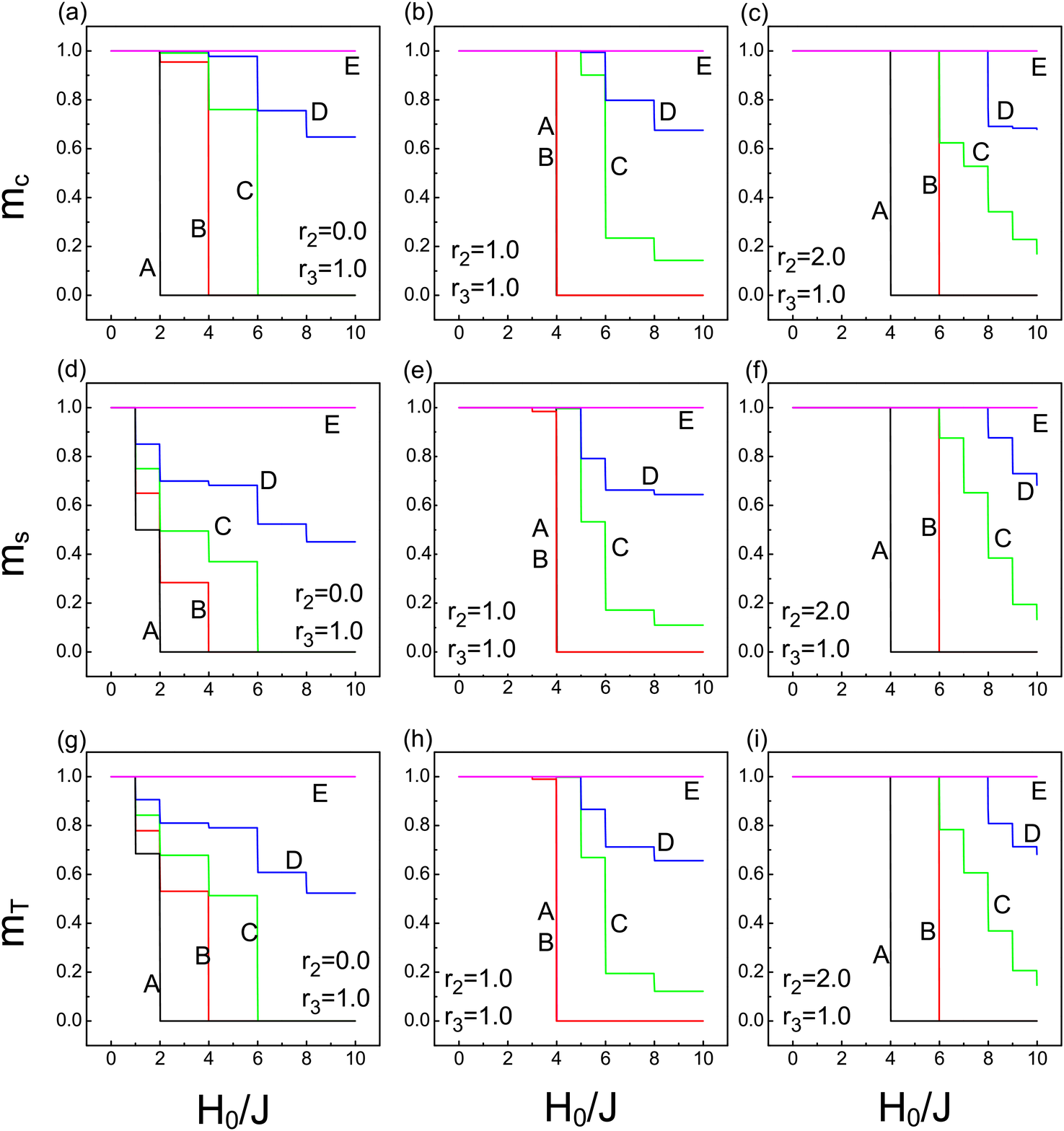, width=14cm,height=14cm}
\end{center}
\caption{Variation of the ground state magnetizations for selected
values of $r_2$ and $p$ for the system with trimodal magnetic field
distribution. Each magnetization calculated at a temperature
$k_BT/J=0.001$ which can be regarded as  ground state and for the
parameter value $r_3=1.0$. The letters which are close to the
related curve represent the different $p$ values as (A) $p=0.0$, (B)
$p=0.3$, (C) $p=0.5$, (D) $p=0.7$, (E) $p=1.0$}
\label{sek8}\end{figure}

The ground state magnetizations appear more complicated forms for
this distribution than the bimodal distribution. This can be seen
from Fig. \ref{sek8} which are corresponding ground state magnetizations to the phase diagrams
given in Fig. \ref{sek6}(a),(c),(e),(g). Construction of the total
ground state magnetization from the ground state magnetizations of
the core and shell is the same as explained in the bimodal
distribution section, since trimodal distribution does not change
the contribution ratios of the core and shell to the total
magnetization. But we can see from Fig. \ref{sek8} that, trimodal
distribution induce higher number of partially ordered phases to the
system both core and shell than the bimodal distribution. This
causes higher number of partially ordered phases of the whole
system. It is not meaningful to give all the ground state
magnetization values of that ordered phases seen in Fig. \ref{sek8},
since that values depends on the parameters $r_2,r_3,p$ and $H_0$. But we
have to say that rising $r_2$ for fixed $r_3=1.0$, shifts the
partially ordered phases to the right of the $m_T-H_0/J$ plane, means
that increasing $r_2$ destroys partially ordered phases with
beginning in low $H_0/J$. We can also see from Fig. \ref{sek8} that the
same effect can be done by rising $p$, i.e. also rising $p$ destroys
the partially ordered phases beginning from the low $H_0/J$. Thus we
can say that rising $r_2$ for other fixed  parameters or  rising $p$
for other fixed  parameters induces same changes on both phase
diagrams and the ground state magnetizations. This similarity is
expected because rising $r_2$ means that interactions between spins
in the shell getting stronger, then after some value of $r_2$ this
interactions can overcome the effect of the random magnetic field
distribution. On the other hand rising $p$ means that the falling
strength of the randomness, then system can escape from that
partially ordered phases which are formed by randomness effects.






\section{Conclusion}\label{conclusion}

The effect of the random magnetic field distribution on the phase
diagrams and ground state magnetizations of the Ising nanowire
investigated. The evolution of the phase diagrams of the system with
bimodal magnetic field distribution to the pure one (i.e. system
with no magnetic field) investigated as the distribution parameter
of the magnetic field goes from $p=0$ to $p=1$. In the bimodal
distribution as a special case of the trimodal distribution with
$p=0$, some partially ordered phases seen for the system as a result
of the inspection of the ground state magnetizations.  This
complicated situation in the ground state magnetizations comes from
the competition between the randomness of the magnetic field which
drag the system to the disordered phase and the interaction
strengths between the spins which try to hold the system in an
ordered phase.

Our findings about the phase diagrams in $(k_BT_c/J-H_0/J)$ plane as follows:
\begin{itemize}

\item All closed phase diagrams intersect $H_0/J$ axes at integer $H_0/J$ values. This means that at zero temperature,
system has ferromagnetic ordering  up to the some value of $H_0$
which is multiple of the core interaction strength $J$.
\item In the absence of the core-shell interaction ($r_3=0.0$) the phase diagrams for different $r_2$ values are the same up to some $r_2<r_2^\star$ value where $r_2^\star>1.0$ for the bimodal distribution. This fact comes from the difference between the total interaction energy of the core and shell. When $r_2$ reaches the value $r_2^\star$, shell
 starts to determine  the critical temperature of the system and thus phase diagrams for $r_2>r_2^\star$ become separate.
\item Phase diagrams of the system with trimodal magnetic field distribution getting open on the high $H_0/J$  region when $p$ is sufficiently large. This is due to the decreasing randomness.
\end{itemize}

Lastly our findings about the ground state magnetization values as follows:

\begin{itemize}
\item System with bimodal magnetic field distribution have a few partially ordered phases.

\item This number of partially ordered phases getting higher in the trimodal distribution.

\item Rising $p$ and also rising $r_2$ or $r_3$ destroys this partially ordered phases.

\end{itemize}

These partially ordered phases requires more detailed investigation
in order to obtain more precise relations of the numerical values of
that ground state magnetizations and the length of a certain ground
state magnetization in the $(m-H_0/J)$ plane, with random magnetic
field distribution parameters. At the same time these partially ordered phases means that, the phase diagrams in $(k_BT_c/j-H_0/J)$ plane
have some first order transition lines. But since one can not calculate the free energy of the system within this formulation, we can not explicitly
determine these first order transition lines.

Altough these questions waiting for
being answered, another important question is the effect of the
continuous random magnetic field distributions (e.g. Gaussian
distribution) to the phase diagrams of the Ising nanowire. This will
be our next work. We hope that the results
 obtained in this work may be beneficial form both theoretical and experimental point of view.

\end{document}